\documentclass[aps,pre,showpacs,superscriptaddress]{revtex4}

\usepackage[dvips]{graphicx}
\usepackage{mathrsfs}
\usepackage{amsmath,amssymb}
\usepackage{hhline}
\usepackage{dcolumn}
\usepackage{bm}

\usepackage[dvips]{color}

\newcommand{\smax}{s_{\rm max}}
\newcommand{\rmax}{r_{\rm max}}
\newcommand{\Smax}{S_{\rm max}}
\newcommand{\Rmax}{R_{\rm max}}
\newcommand{\ps}{\psi_{\rm S}}
\newcommand{\pr}{\psi_{\rm R}}
\newcommand{\lsir}{\lambda_c^{\rm SIR}}
\newcommand{\lcl}{\lambda_{c1}}
\newcommand{\lcu}{\lambda_{c2}}

\begin{document}

\title{Outbreaks in susceptible-infected-removed epidemics with multiple seeds}

\author{Takehisa Hasegawa}
\email{takehisa.hasegawa.sci@vc.ibaraki.ac.jp}
\affiliation{%
Department of Mathematics and Informatics, 
Ibaraki University, 
2-1-1, Bunkyo, Mito, 310-8512, Japan
}%
\author{Koji Nemoto}
\email{nemoto@statphys.sci.hokudai.ac.jp}
\affiliation{%
Department of Physics, Hokkaido University,
Kita 10 Nishi 8, Kita-ku, Sapporo, 060-0810, Japan
}%

\begin{abstract}
We study a susceptible-infected-removed (SIR) model with multiple seeds on a regular random graph. 
Many researchers have studied the epidemic threshold of epidemic models above which a global outbreak can occur, starting from an infinitesimal fraction of seeds. 
However, there have been few studies of epidemic models with finite fractions of seeds. 
The aim of this paper is to clarify what happens in phase transitions in such cases.
The SIR model in networks exhibits two percolation transitions.
We derive the percolation transition points for the SIR model with multiple seeds 
to show that as the infection rate increases epidemic clusters generated from each seed percolate before a single seed can induce a global outbreak.
\end{abstract}


\maketitle


\section{Introduction}

The threat of infectious disease is becoming increasingly conspicuous for modern society, 
wherein there is a large amount of international travel all over the world.
Understanding how infectious diseases spread in our society is crucial to the development of strategies for disease control. 
A mathematical model of infectious disease, called the susceptible-infected-removed (SIR) model, was first applied with the assumption of a well-mixed population 
for computation of the final numbers of infected and eventually removed (or recovered) individuals \cite{kermack1927contribution}.
So far, many mathematical models of infectious diseases have been proposed for understanding the spread of epidemics 
and proposing strategies for disease control \cite{anderson1992infectious}.

In recent years, many studies have been devoted to epidemic models with a network structure of people \cite{pastor2014epidemic}. 
Diseases spread over the networks of physical contacts between individuals, and the structure of real networks \cite{albert2002statistical,newman2003structure,dorogovtsev2008critical,barrat2008dynamical} has crucial effects on this spread.
For example, Moreno et al. \cite{moreno2002epidemic} studied the SIR model in a scale-free network 
having a degree distribution of $p_k \propto k^{-\gamma}$ using a degree-based mean-field approach.
Their approximation clarified that epidemics can spread over the network for any infection rate if $\gamma \le 3$.
In addition, many analytical approaches for epidemic models with network structures, 
such as the approximation onto a bond percolation problem \cite{grassberger1983critical,newman2002spread}, 
the edge-based compartment model \cite{miller2011edge}, the effective degree approach \cite{lindquist2011effective}, 
and the pair approximation \cite{gleeson2011high,gleeson2013binary},
have been proposed and have succeeded in describing epidemic dynamics.
Numerical simulations have revealed how epidemics spread in more realistic situations. 
Also, several strategies for disease control have been proposed on the basis of the knowledge of epidemics on networks, 
e.g., target immunization \cite{pastor2002immunization, madar2004immunization}, 
acquaintance immunization \cite{madar2004immunization, gallos2007improving, cohen2003efficient, holme2004efficient}, 
and graph-partitioning immunization \cite{chen2008finding}.

Most previous studies using SIR-type epidemic models have assumed that the fraction of infection seeds is infinitesimally small. 
In contrast, there have been few studies on epidemic models with finite fractions of seeds. 
Miller \cite{miller2014epidemics} considered the SIR model in networks with large initial conditions 
to resolve an apparent paradox in works assuming an infinitesimal fraction of seeds. 
Hu et al. \cite{hu2014effects} numerically studied how the positions of multiple seeds in a network affect spreading behavior. 
Ji et al. \cite{ji2015effective} identified multiple influential spreaders in real networks by ranking nodes in disintegrated networks after random bond removals.
What we discuss here is a more fundamental, but almost overlooked problem: 
How do epidemic models with finite fractions of seeds undergo phase transitions?
For SIR-type epidemics, each infection seed creates an epidemic cluster of infected individuals. 
Epidemic clusters generated by multiple seeds will have global connectivity in some parameter regions 
even though each seed may not have the potential to induce a global outbreak there.

In this paper, we consider the SIR model in networks with multiple seeds.
In this case, the SIR model exhibits a kind of percolation transition.
An epidemic cluster grows from each of multiple seeds.
We regard the clusters so generated as {\it supernodes} and study the percolation problem of these supernodes. 
Indeed, we can analytically and numerically obtain the percolation transition point of supernodes to show a gap between this transition point and the epidemic threshold.
The existence of this gap indicates that the percolation transition of epidemic clusters occurs before a single seed can induce a global outbreak. 
Our result also shows the sensitivity of the seed fraction on percolation transition points,
i.e., that a small seed fraction drastically reduces the critical infection rate for the emergence of the infinite epidemic cluster.

\section{Model}

Let us give a brief review of the SIR model in a given static network. 
Each node in the network takes one of three states: susceptible, infected, and removed. 
The system evolves as a continuous-time Markov process.
As an initial state configuration, a fraction, $\rho$, of the nodes is randomly chosen to be seeds and is initially infected, while other nodes are susceptible.
The infection rate is denoted by $\lambda$.
When an infected node is adjacent to a susceptible node, this susceptible node gets infected with probability $\lambda \Delta t$ within a short time, $\Delta t$.
Note that this probability is independently given by each of the infected nodes so that 
the total infection rate at a node is just proportional to the number of infected neighbors.
An infected node becomes removed at a rate $\mu$,
i.e., with probability $\mu \Delta t$ within a short time $\Delta t$, irrespective of the neighbors' states. 
Without loss of generality, we set $\mu=1$ unless otherwise specified. 
The dynamics stops when no infected nodes exist in the network.

Let us consider the limit $\rho \to 0$.
The SIR model exhibits a phase transition at the epidemic threshold $\lambda=\lsir$ when $\lambda$ increases from zero.
Above $\lsir$, a single seed can induce global outbreaks.
In a global outbreak, a nonzero fraction of nodes become infected and eventually removed.
Below $\lsir$, the number of removed nodes is always negligible compared with the total number of nodes.
As already mentioned, we have several approaches for obtaining $\lsir$ (see the recent review \cite{pastor2014epidemic});
Newman approximated the SIR model on uncorrelated networks by mapping onto a bond percolation problem
(which is called the SIR model with transmissibility) \cite{newman2002spread} and derived $\lsir$ as 
\begin{equation}
\lsir = \frac{\langle k \rangle}{\langle k^2 \rangle-2\langle k \rangle}, \label{lambda_sir}
\end{equation}
where $\langle k \rangle$ and $\langle k^2 \rangle$ are the first and the second moments of the degree distribution, $p_k$, respectively 
\footnote{
Here we replaced the transmissibility $T$ in \cite{newman2002spread} with $\lambda$ 
as $T=\lambda/(\mu+\lambda)$ with $\mu=1$.
In \cite{newman2002spread}, transmissibility $T$ is given as 
$T=\langle T_{ij} \rangle=1-\int_{0}^\infty {\rm d}r {\rm d}\tau P(r) P(\tau) e^{-r \tau}$, 
where $T_{ij}$ is the probability that the disease did not transmit from node $i$ to node $j$ 
before node $i$ was removed when we consider the pair of an infected node $i$ and a susceptible node $j$, 
$r$ is the infection rate between the focal pair, $\tau$ is the time for which the infected node remains infected, 
and $P(r)$ and $P(\tau)$ are the corresponding distributions.
In the present case, we have $P(r)=\delta(r-\lambda)$ (because the infection rate is a constant $\lambda$), 
$P(\tau)=\mu e^{-\mu \tau}$ (as the distribution of the inter-event time of a Poisson process with parameter $\mu$), 
and have $T=1-\int_{0}^\infty {\rm d}r {\rm d}\tau \delta(r-\lambda) \mu e^{-\mu \tau} e^{-r \tau}
=1-\mu \int_{0}^\infty {\rm d}\tau e^{-(\mu+\lambda) \tau}=\lambda/(\mu+\lambda)$.
}.
This result indicates that, for a fat-tailed scale-free network whose degree distribution obeys $p_k \propto k^{-\gamma}$with $\gamma \le 3$, 
a global outbreak starting from an infinitesimal fraction of seeds occurs even for an infinitesimal infection rate.
As indicated in \cite{kenah2011epidemic}, mapping onto a bond percolation problem does not give the exact outbreak size or probability, 
but it does predict exactly the epidemic threshold. 
Lindquist et al. \cite{lindquist2011effective} proposed an effective degree approach for describing 
the time evolution of the SIR dynamics using numerous ordinary differential equations and derived the same epidemic threshold (\ref{lambda_sir}). 
Miller \cite{miller2011edge} introduced another approach by means of the edge-based compartment model 
to enable accurate descriptions of the SIR dynamics accurately with a few rate equations.

We can also describe the phase transition of the present model in terms of percolation.
In any final state, each node takes either a susceptible or a removed state.
We call the connected components of removed nodes and susceptible nodes the R-components and the S-components, respectively.
For the SIR model on networks with $\rho \gtrsim 0$, we have two percolation transition points, $\lcl$ and $\lcu$. 
When the number of nodes, $N$, is much greater than one, 
the mean fraction of the largest R-component, $\rmax(N)=\Rmax(N)/N$, where $\Rmax(N)$ is the mean size of the largest R-component, 
changes from zero to a nonzero value at the former point $\lcl$. 
Note that $\lcl$ corresponds to $\lsir$ in the limit $\rho \to 0$ by definition. 
Percolation analysis of an epidemic cluster starting from a single seed has been used for numerical computations 
of the epidemic threshold and critical properties \cite{tome2010critical, de2011new}. 
The latter point, $\lcu$, is on the percolation of the S-component (also called the residual graph \cite{newman2005threshold,ferrari2006network}) 
and is usually larger than $\lcl$. 
Above $\lcu$, the network remaining after removal of the R-components is disintegrated such that the sizes of all remaining components are finite.
In other words, the mean fraction of the largest S-component, $\smax(N)=\Smax(N)/N$, where $\Smax(N)$ is the mean largest S-component size, 
is zero (nonzero) for $\lambda>\lcu$ ($\lambda<\lcu$) when $N \gg 1$.
Whether the susceptible nodes are globally connected is important because a second epidemic spread may occur in the remaining network \cite{newman2005threshold,bansal2012impact}. 
In \cite{newman2005threshold}, Newman analyzed this second transition point of the SIR model with transmissibility in uncorrelated networks with $\rho \to 0$ 
to show that the transition point is positive even when $\gamma \le 3$.
Valdez et al. \cite{valdez2012intermittent} proposed a new strategy for suppressing epidemics 
by regarding this second transition point as a measure of the efficiency of a mitigation or control strategy. 
If we regard the present model as showing the propagation of an attack against a network, such as a computer virus, 
$\lcu$ is a measure of the robustness of networks against such attacks \cite{hasegawa2011robustness,hasegawa2012robustness}.
Konno and the authors numerically studied $\lcu$ for correlated networks to show that 
any positive or negative degree correlation makes networks more robust \cite{hasegawa2012robustness}.

To summarize, the system with a given value of $\rho$ has the following three regions: 
(i) the S-dominant phase, where $\rmax =0$ and $\smax >0$ for $\lambda<\lcl$; 
(ii) the coexisting phase, where $\rmax >0$ and $\smax >0$ for $\lcl<\lambda<\lcu$; 
and (iii) the R-dominant phase, where $\rmax >0$ and $\smax =0$ for $\lambda>\lcu$. 
To investigate in detail the phase transitions of the SIR model with a finite fraction of seeds, 
we focus on the $z$-regular random graph (RRG). 
Our formulations discussed below are for the RRG. 
The extension to degree-uncorrelated networks having degree distribution $p(k)$ may be straightforward, 
although its execution will be cumbersome. 
At any rate, our findings obtained from the RRG probably will be in common with other networks.
In Sec.~\ref{spatial}, 
we numerically study the outbreaks induced by multiple seeds in finite dimensional Euclidean lattices.


\begin{figure}
 \begin{center}
  \includegraphics[width=75mm]{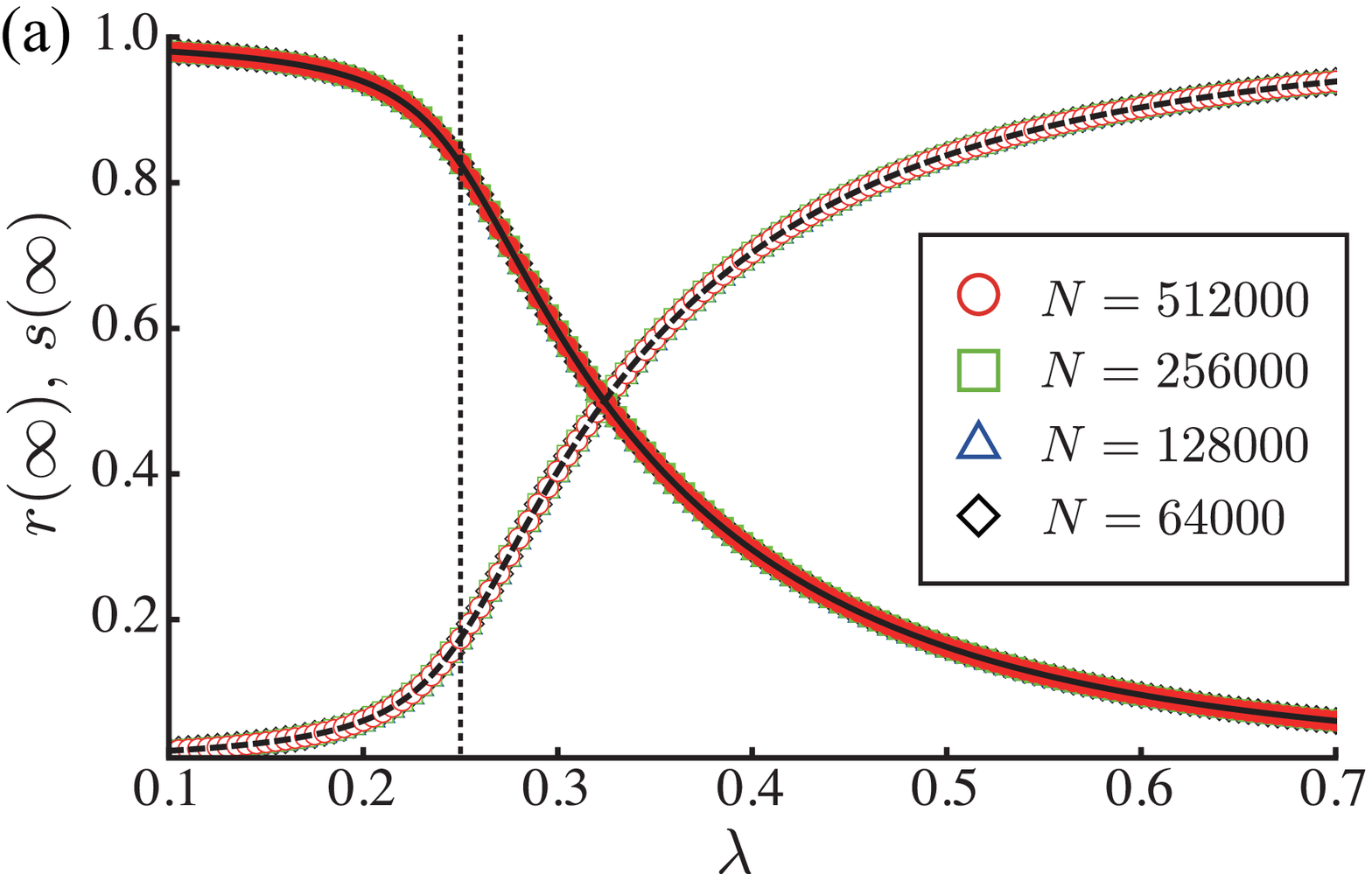}
  \includegraphics[width=75mm]{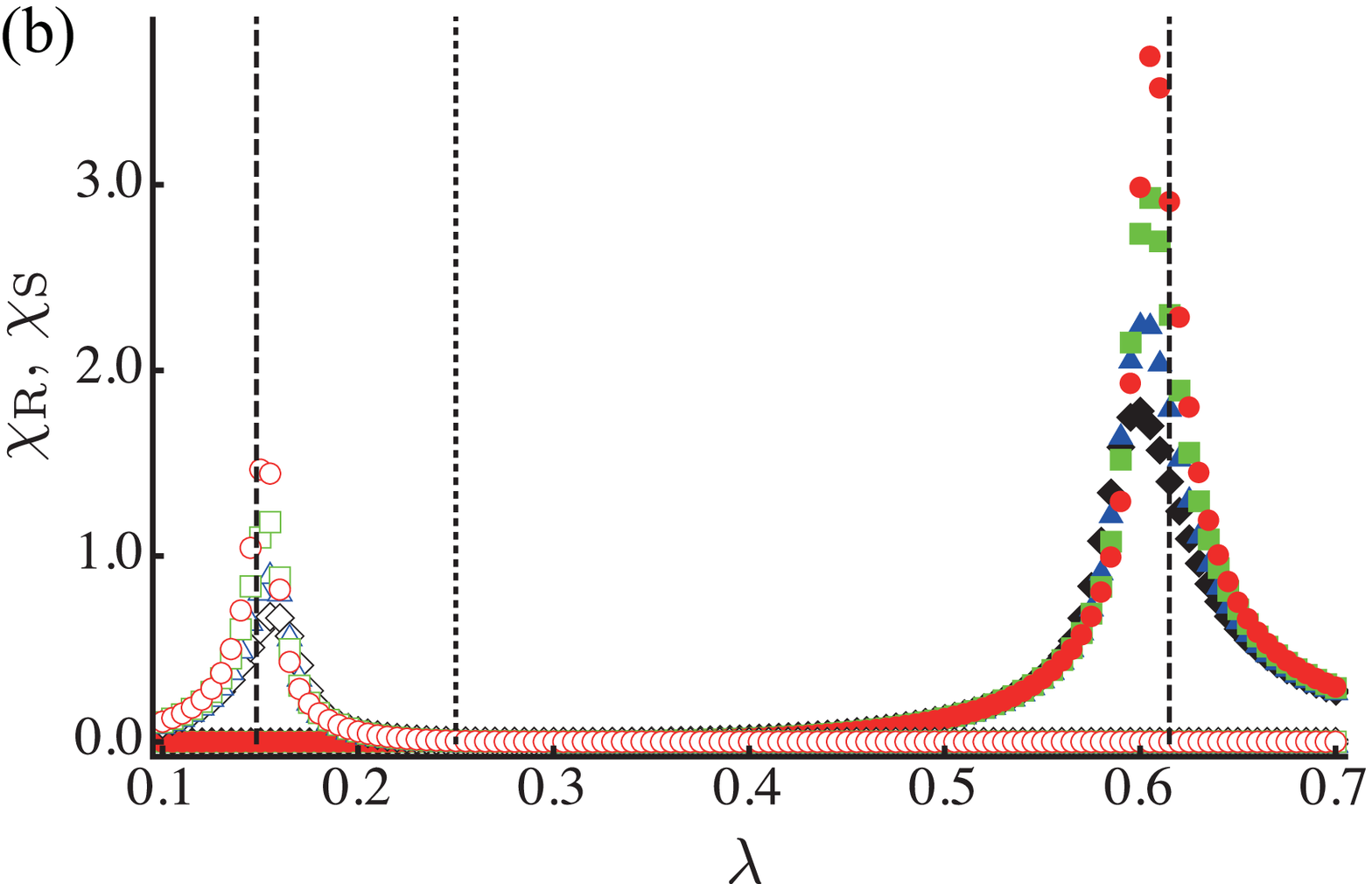}
   \end{center}
 \caption{
Numerical results for the SIR model with $\rho=0.01$ on the RRG: 
(a) total densities of the removed nodes $r$ (open symbols) and of the susceptible nodes $s$ (filled symbols) 
and (b) susceptibility of the R-components $\chi_{\rm R}$ (open symbols) and of the S-components $\chi_{\rm S}$ (filled symbols).
The dashed and solid lines in (a) are drawn from the AMEs. 
In (b), the dotted vertical line represents $\lsir$, and the two dashed vertical lines represent 
$\lcl = 0.148$ and $\lcu = 0.615$, which are given in Sec.\ref{sec:multiple}.
}
 \label{fig-AME}
\end{figure}

\section{Total densities of susceptible and removed nodes \label{AME}}

To evaluate the time evolution of the SIR dynamics and the total densities of the susceptible and removed nodes in the final states, 
we consider the approximate master equations (AMEs) \cite{lindquist2011effective, gleeson2013binary}.

Let $s_{l,m}(t)$, $i_{l,m}(t)$, and $r_{l,m}(t)$ be the fractions of nodes that are susceptible, infected, and removed, respectively, 
at time $t$ and have $l$ susceptible and $m$ infected neighbors. 
The AMEs for the evolution of these variables are as follows (see \cite{lindquist2011effective} for details):
 \begin{equation}
 \dot s_{l,m} = -\lambda m s_{l,m} + \beta_s^{si}[(l+1)s_{l+1,m-1}-ls_{l,m}] + \beta_s^{ir}[(m+1)s_{l,m+1}-ms_{l,m}], \label{eq:AMEs}
 \end{equation}
 \begin{equation}
 \dot i_{l,m} = \lambda m s_{l,m} -\mu i_{l,m}+ \beta_i^{si}[(l+1)i_{l+1,m-1}-li_{l,m}]+ \beta_i^{ir}[(m+1)i_{l,m+1}-mi_{l,m}], \label{eq:AMEi}
 \end{equation} 
 \begin{equation}
 \dot r_{l,m} = \mu i_{l,m}+ \beta_r^{si}[(l+1)r_{l+1,m-1}-lr_{l,m}]+ \beta_r^{ir}[(m+1)r_{l,m+1}-mr_{l,m}]. \label{eq:AMEr}
 \end{equation} 
The transition rates of neighboring nodes are approximated as
 \begin{equation}
 \beta_s^{si}=\lambda\frac{\sum_{l,m}lms_{l,m}}{\sum_{l,m}ls_{l,m}},\quad
 \beta_s^{ir}=\mu\frac{\sum_{l,m}li_{l,m}}{\sum_{l,m}li_{l,m}}=\mu, \label{eq:AMEsRate}
 \end{equation}
 \begin{equation}
 \beta_i^{si}=\lambda\frac{\sum_{l,m}m^2s_{l,m}}{\sum_{l,m}ms_{l,m}},\quad
 \beta_i^{ir}=\mu\frac{\sum_{l,m}mi_{l,m}}{\sum_{l,m}mi_{l,m}}=\mu, \label{eq:AMEiRate}
 \end{equation}
 \begin{equation}
 \beta_r^{si}=\lambda\frac{\sum_{l,m}(k-l-m)ms_{l,m}}{\sum_{l,m}(k-l-m)s_{l,m}},\quad
 \beta_r^{ir}=\mu, \label{eq:AMErRate}
 \end{equation}
where the summations run over all $0\le l+m\le k$. 
To describe the SIR dynamics with $\rho>0$, we set the initial condition as
\begin{align}
s_{l,m}(0)&=\delta_{k,l+m}(1-\rho){k \choose l}(1-\rho)^l\rho^m,\\
i_{l,m}(0)&=\delta_{k,l+m}\rho{k \choose l}(1-\rho)^l\rho^m,\\
r_{l,m}(0)&=0. \label{eq:AMEdescribe}
\end{align}

By numerical evaluation of the above equations, we obtain the total densities 
\begin{equation}
s(t)=\sum_{l,m}s_{l,m}(t),\quad
i(t)=\sum_{l,m}i_{l,m}(t),\quad
r(t)=\sum_{l,m}r_{l,m}(t), \label{eq:AMEtotal}
\end{equation}
which satisfy the conservation law, 
\begin{equation}
s(t) + i(t) + r(t) = 1, 
\end{equation}
at any time $t$.
Note that all variables other than $s_{l,0}$ and $r_{l,0}$ vanish in the limit $t \to \infty$, and therefore $i(\infty)=0$.

To check the accuracy of the AME, we perform Monte-Carlo simulations for the SIR model on the RRG with $z=6$. 
In our simulations, we set $\mu=1$ and $\rho=0.01$. 
The numbers of nodes are $N=64000, 128000, 256000$, and $512000$. 
The number of graph realizations is 100, and the number of trials on each graph is 500.
Figure~\ref{fig-AME}(a) shows the AME result (line) and the Monte-Carlo result (symbols) of 
the total densities of susceptible and removed nodes, $s$ and $r$, in the final states.
We find that data from the AMEs wholly coincides with those from the Monte-Carlo simulations.

Equations (\ref{eq:AMEs}-\ref{eq:AMEtotal}) do not predict any transition point for $\rho>0$ because $r(\infty) \ge \rho>0$, 
although it is possible to derive the epidemic threshold $\lsir=\mu/(z-2)$ for the RRG with degree $z$ \cite{lindquist2011effective} by considering the limit $\rho \to 0$ (see Appendix \ref{sec:derivation-lsir}).
In contrast, the Monte-Carlo simulations suggest that the model actually exhibits phase transitions. 
In Fig.~\ref{fig-AME} (b), we plot the R- and S-susceptibilities 
(which we call the susceptibility by analogy with the magnetic susceptibility in spin systems) $\chi_{\rm R}$ and $\chi_{\rm S}$. 
Here, $\chi_{\rm R}$ ($\chi_{\rm S}$) is defined as the mean size of all R-components (all S-components) except the largest one \footnote{
The susceptibilities $\chi_{\rm R}$ and $\chi_{\rm S}$ are given as
$\chi_{\rm R} = \sum_{r \neq r_{\rm max}} r^2 n_{\rm R}(r)$ and 
$\chi_{\rm S} = \sum_{s \neq s_{\rm max}} s^2 n_{\rm S}(s)$, where $n_{\rm R}(r)$ ($n_{\rm S}(s)$) means the mean number of R-components (S-components) of size $r$ ($s$) per node.}. 
We find that $\chi_{\rm R}$ and $\chi_{\rm S}$ have peaks at $\lcl$ and $\lcu$, respectively, implying two phase transitions. 
Moreover, these points are clearly different from $\lsir$. 
In particular, the gap between $\lcl$ and $\lsir$ indicates that as the infection rate increases, 
the epidemic clusters generated from each seed percolate before a single seed can induce a global outbreak.
In the next section, we derive these percolation transition points for $0<\rho<1$.


\section{Percolation transitions of the SIR dynamics with multiple seeds \label{sec:multiple}}

\begin{figure}
 \begin{center}
  \includegraphics[width=55mm]{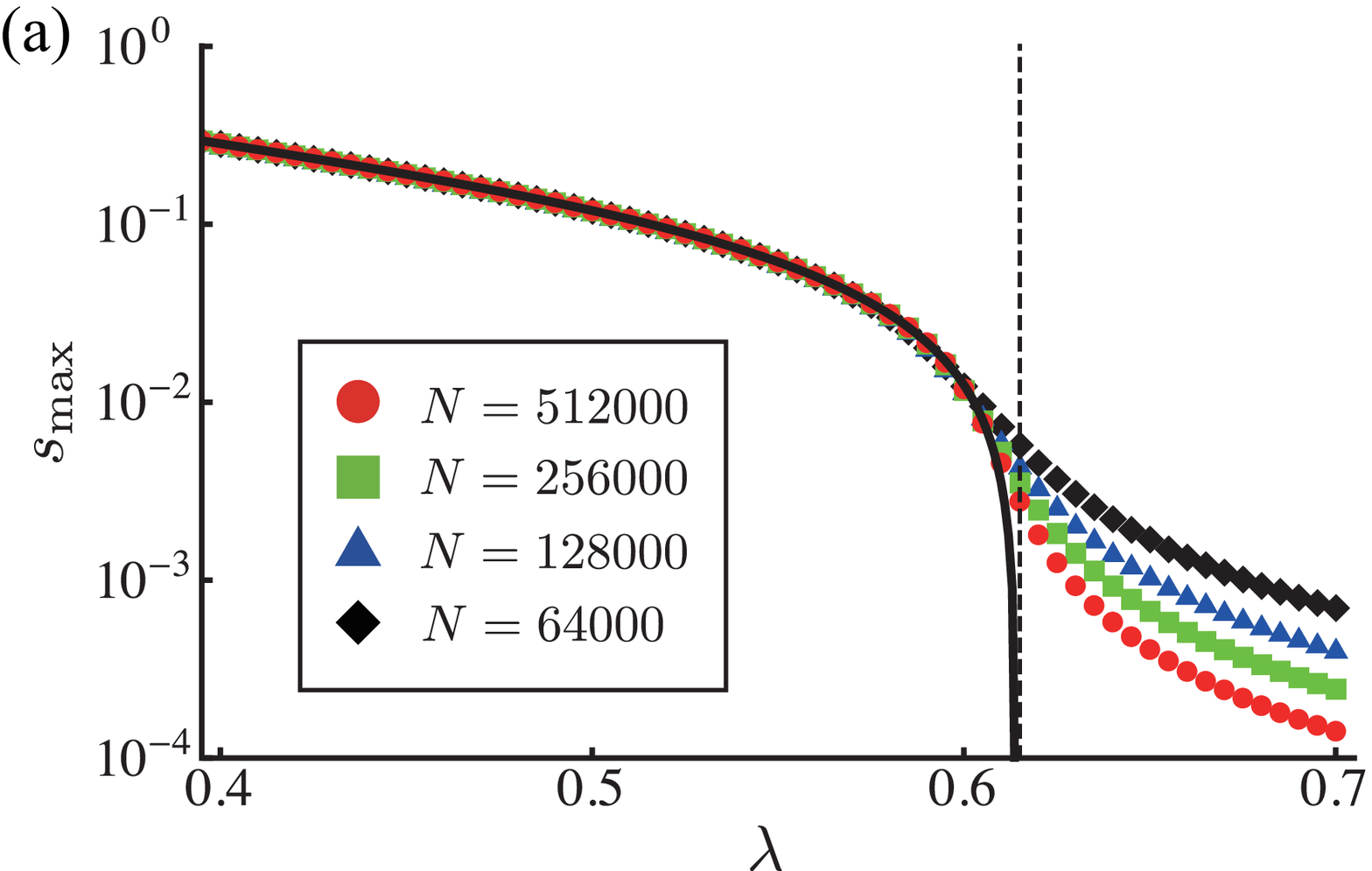}
  \includegraphics[width=55mm]{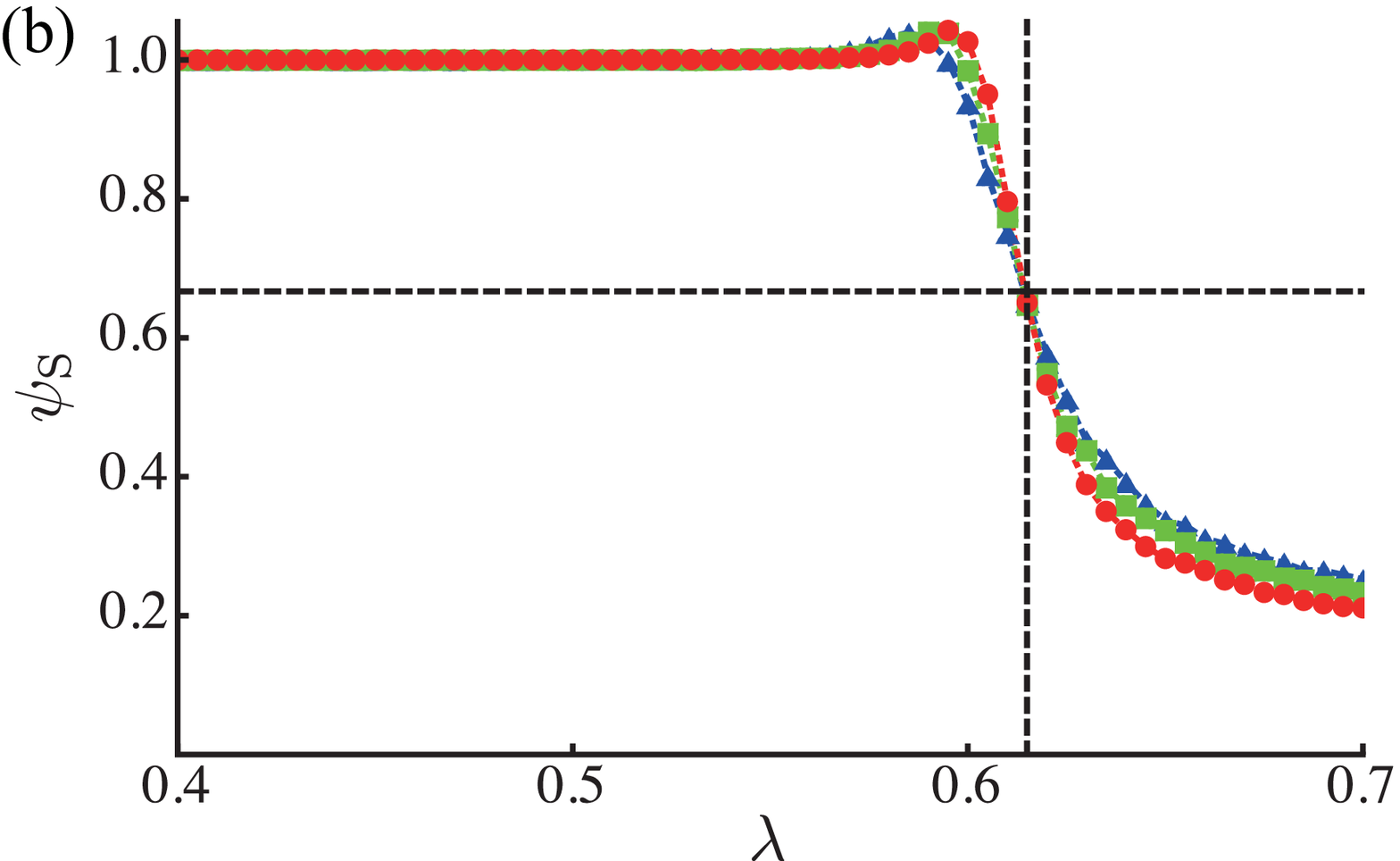}
  \includegraphics[width=55mm]{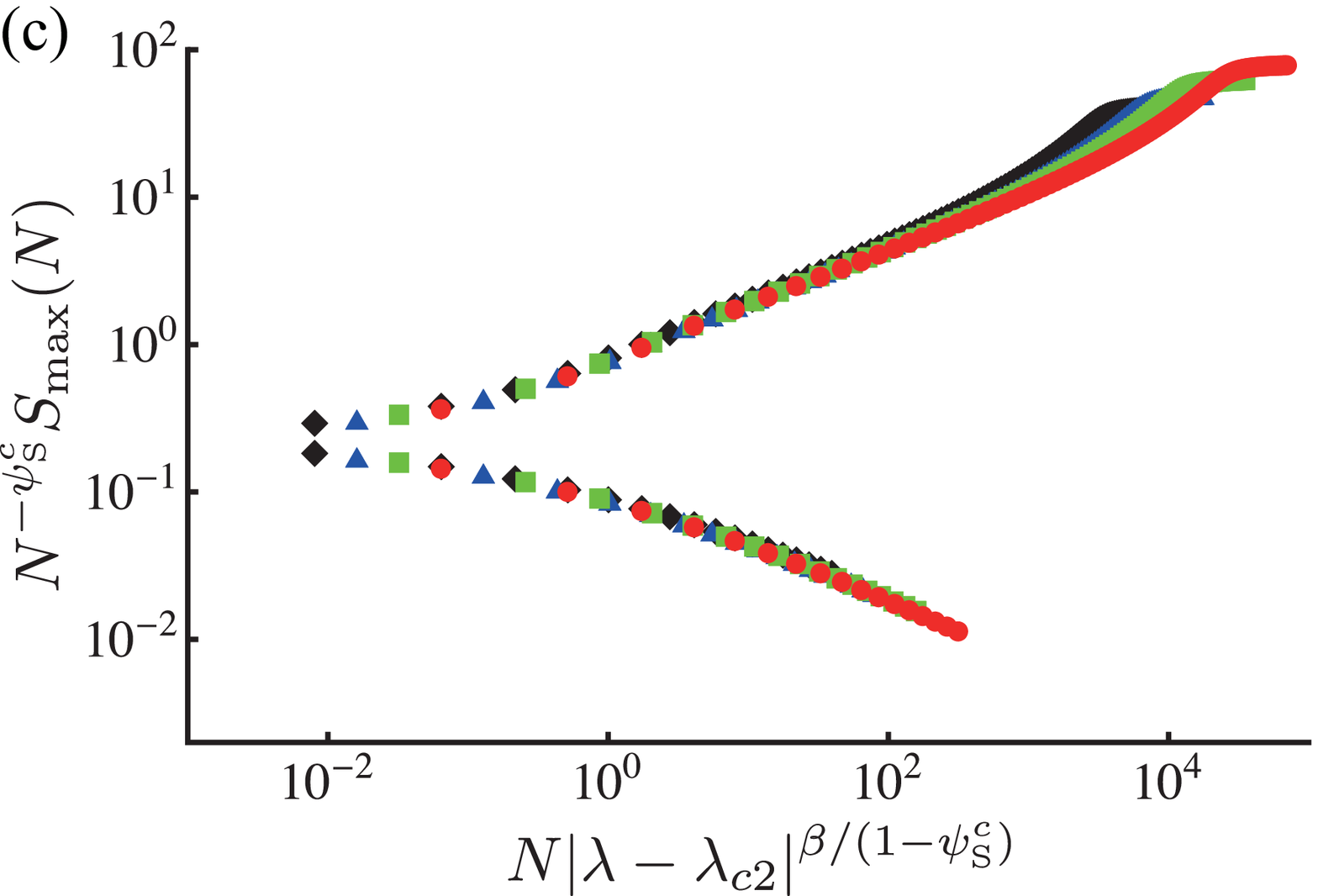}
   \end{center}
 \caption{
 (a) Fraction of the largest S-component, $s_{\rm max}$, as a function of $\lambda$.
 Symbols were obtained from Monte-Carlo simulations.
 The solid line was drawn by evaluating Eq.~(\ref{smax}).
 (b) Numerical results of the fractal exponent of the largest S-component, $\ps$.
 Data with several values of $N$ have a crossing point at $\lcu \simeq 0.615$ and $\psi_{\rm S}^c \simeq 2/3$.
 (c) Finite size scaling for $S_{\rm max}$ around $\lcu$.
 Here we set $\psi_{\rm S}^c=2/3$ and $\beta=1$, 
 which means that the transition at $\lcu$ belongs to the mean-field universality class.
}
 \label{fig:smax}
\end{figure}

\subsection{Derivation of $\lcu$}

To derive $\lcu$, we consider the percolation of the S-components. 
In \cite{newman2005threshold}, Newman analyzed the percolation of the S-components using generating functions. 
His method gives $\lcu$ for the SIR model in uncorrelated networks but assumes a single seed. 
By combining the AMEs and Newman's method, we obtain $\smax$ and $\lcu$ for the case with $0<\rho<1$.

Let us consider the S-components in a typical final state for the SIR model on an infinitely large RRG with $\rho>0$. 
In the previous section, we already have the probability $s_{l,0}(\infty)$ that a randomly chosen node is susceptible and has $l$ susceptible neighbors ($s_{l,m}(\infty)=0$ for $m \neq 0$).
Using $s_{l,0}(\infty)$, we obtain the degree distribution of the S-components as
\begin{equation}
p^s_l=\frac{1}{s} s_{l,0}(\infty), 
\end{equation}
where the denominator $s= \sum_{l=0}^k s_{l,0}$ is the prior probability of being susceptible. 
The corresponding generating function, $F^s_0(x)$, is given by 
\begin{equation}
F^s_0(x)=\sum_{l=0}^k p^s_l x^l.
\end{equation}
Here we assume that this subnetwork is degree-uncorrelated. 
We consider the excess degree, which is the degree of the node reached by following a randomly chosen link minus one 
\cite{newman2003structure}. 
The excess degree distribution, $q_l^s$, which is the probability that a randomly chosen link from S-components points 
to a (susceptible) node with excess degree $l$, is $(l+1) p_{l+1}^s/\sum_{l'} (l'+1) p_{l'+1}^s$.
Then the generating function for the excess degree distribution of the S-components is 
\begin{equation}
F^s_1(x)=\frac{{F^s_0}'(x)}{{F^s_0}'(1)},
\end{equation}
and the mean excess degree is given by ${F^s_1}'(1)$. 
By arguing the emergence of an infinitely connected component of this subnetwork (similar to \cite{newman2003structure}), 
we easily find that there is an infinite S-component if ${F^s_1}'(1)>1$, and thus, the percolation transition point of the S-component, $\lcu$, satisfies 
\begin{equation}
{F^s_1}'(1)=1.
\end{equation}

Following \cite{newman2005threshold}, we also have the mean fraction of the largest S-component, $\smax$, as 
\begin{equation}
\smax=s[1-F^s_0(v)], \label{smax}
\end{equation}
where $v$ is the solution of
\begin{equation}
v=F^s_1(v).
\end{equation}

We check these estimates using Monte-Carlo simulations. 
Figure~\ref{fig:smax}(a) shows the order parameter, i.e., the fraction of the largest S-component, $s_{\rm max}(N)$, for RRGs with several $N$'s.
We find that the numerical results coincide with the analytical line below $\lcu$ and tend to zero with increasing $N$ above $\lcu$. 
To numerically obtain the transition point, $\lcu$, we introduce the fractal exponent \cite{hasegawa2013profile}. 
The fractal exponent of the largest S-component is defined and approximated as
\begin{equation}
\ps=\frac{{\rm d} \ln \Smax(N)}{{\rm d} \ln N} \approx \frac{\ln \Smax(N)-\ln \Smax(N/2)}{\ln N- \ln (N/2)}. 
\end{equation} 
In the limit $N \to \infty$, $\ps=1$ for $\lambda<\lcu$ and $\ps=0$ for $\lambda>\lcu$, 
because the largest S-component size should be proportional to $N$ for $\lambda<\lcu$ and finite for $\lambda>\lcu$.
As shown in Fig.~\ref{fig:smax}(b), numerical results for $\ps$ approach $\ps=1 (\ps=0)$ for $\lambda<\lcu$ ($\lambda>\lcu$) as $N$ increases, and have a crossing point at $\lcu$.
From numerical data, we have $\ps^c \equiv \ps(\lcu) \simeq 2/3$ at $\lcu \simeq 0.615$, 
which coincides well with our analytical estimate (vertical line in Fig.~\ref{fig:smax}).

This observation is also confirmed by a finite size scaling. 
As in \cite{hasegawa2013profile}, we assume a scaling form for $\Smax(N)$ as
\begin{eqnarray}
\Smax(\lambda, N) &=& N^{\ps^c} f_s [ N (\Delta \lambda)^{\beta/(1-\ps^c)}],
\label{eq:smax_scaling}
\end{eqnarray}
where $\Delta \lambda=|\lcu-\lambda|$, $\beta$ is the critical exponent related to the order parameter, $\smax \propto \Delta \lambda^\beta$, and 
\begin{equation}
f_s(x) \propto 
{\Biggl\{}
\begin{array}{ccl}
{\rm const} & {\rm for} & x \ll 1, \\
x^{1-\ps^c} & {\rm for} & x \gg 1.
\end{array} \label{eq:sscaling_func}
\end{equation}
In Fig.~\ref{fig:smax}(c), our scaling shows a nice collapse with $\ps^c=2/3$ and $\beta=1$. 
Because $\ps^c$ is related to another critical exponent $\tau$ as $\ps^c=(\tau-1)^{-1}$\cite{hasegawa2013profile}, where $\tau$ is associated with the distribution function of S-components 
$n_{\rm S}(s)$ at the critical point, $n_{\rm S}(s) \propto s^{-\tau}$, 
these exponents mean that the percolation transition of the S-components actually occurs at $\lcu$ and belongs to the mean-field universality class such that $\beta=1$ and $\tau=5/2$ \cite{stauffer1994introduction}.

\begin{figure}
 \begin{center}
  \includegraphics[width=90mm]{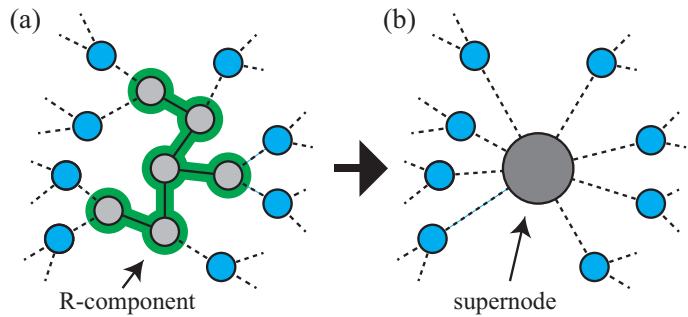}
   \end{center}
 \caption{
Example of (a) an R-component and (b) the corresponding supernode. 
In this example, the R-component, of size $n=6$ on a 3-RRG, has $n-1=5$ edges inside (solid lines) 
and $z n - 2(n-1)=8$ edges attached to susceptible nodes (dashed lines). 
Thus the degree of the supernode is $k_n=8$.
}
 \label{fig:super}
\end{figure}

\begin{figure}
 \begin{center}
  \includegraphics[width=55mm]{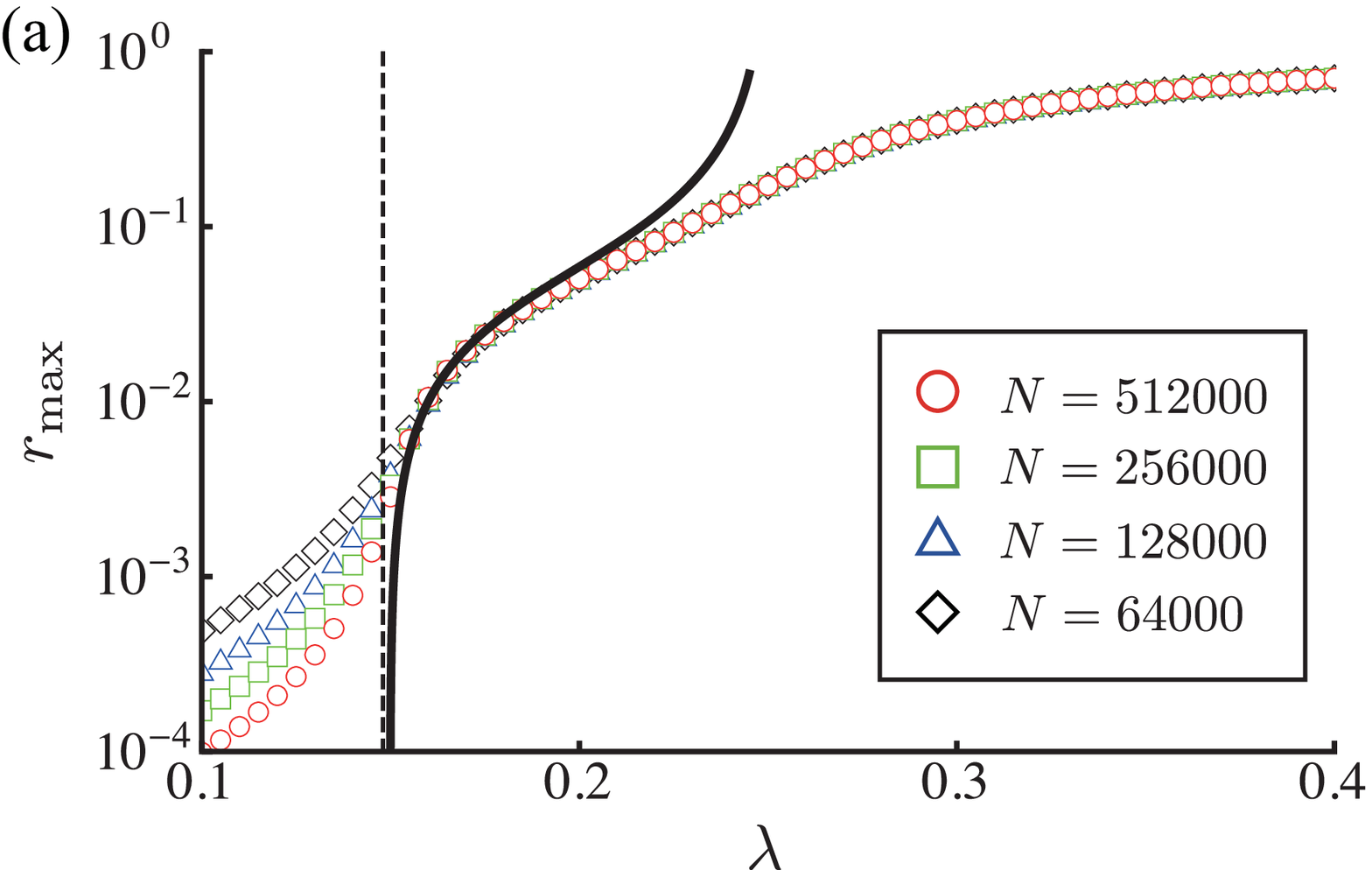}
  \includegraphics[width=55mm]{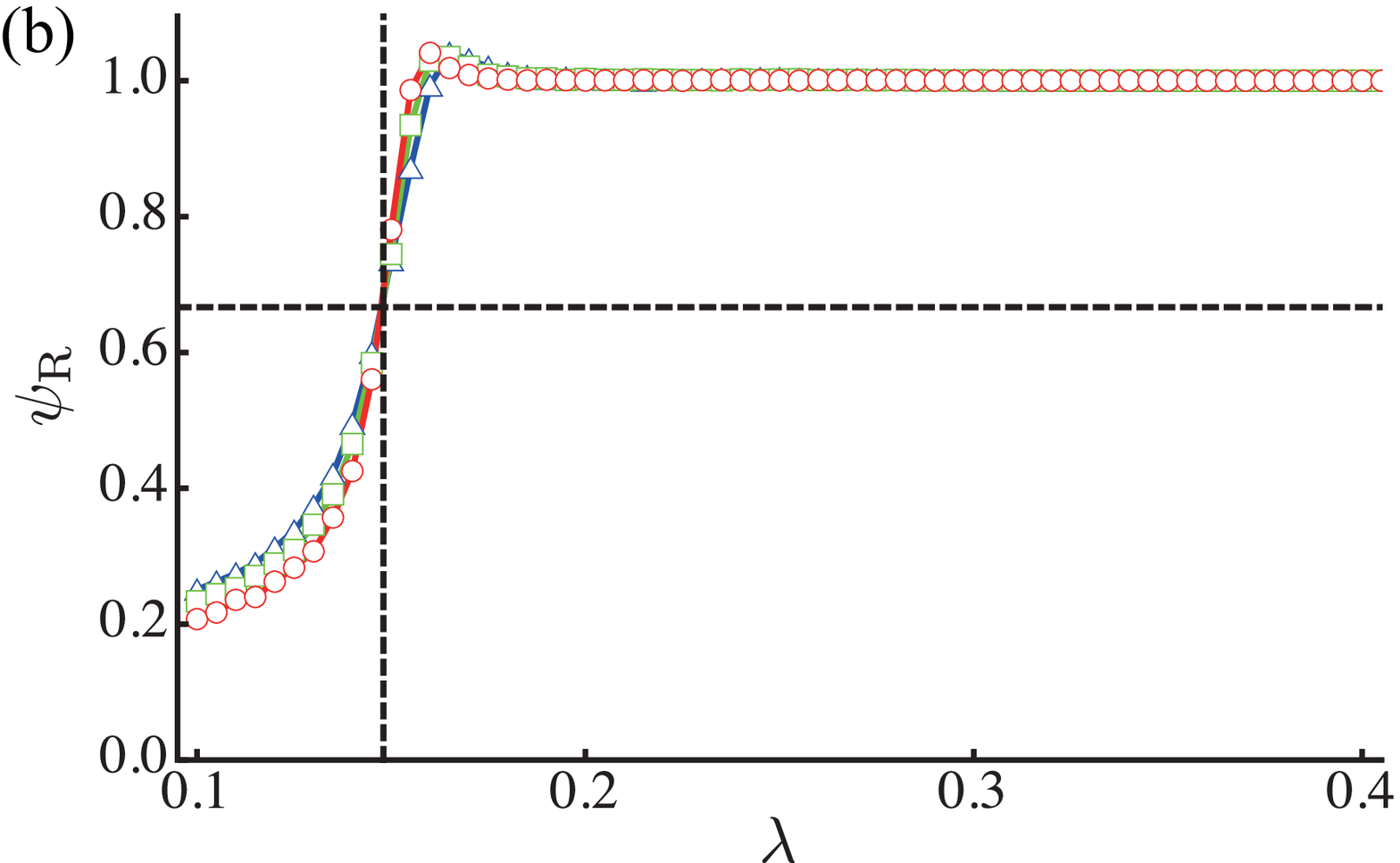}
  \includegraphics[width=55mm]{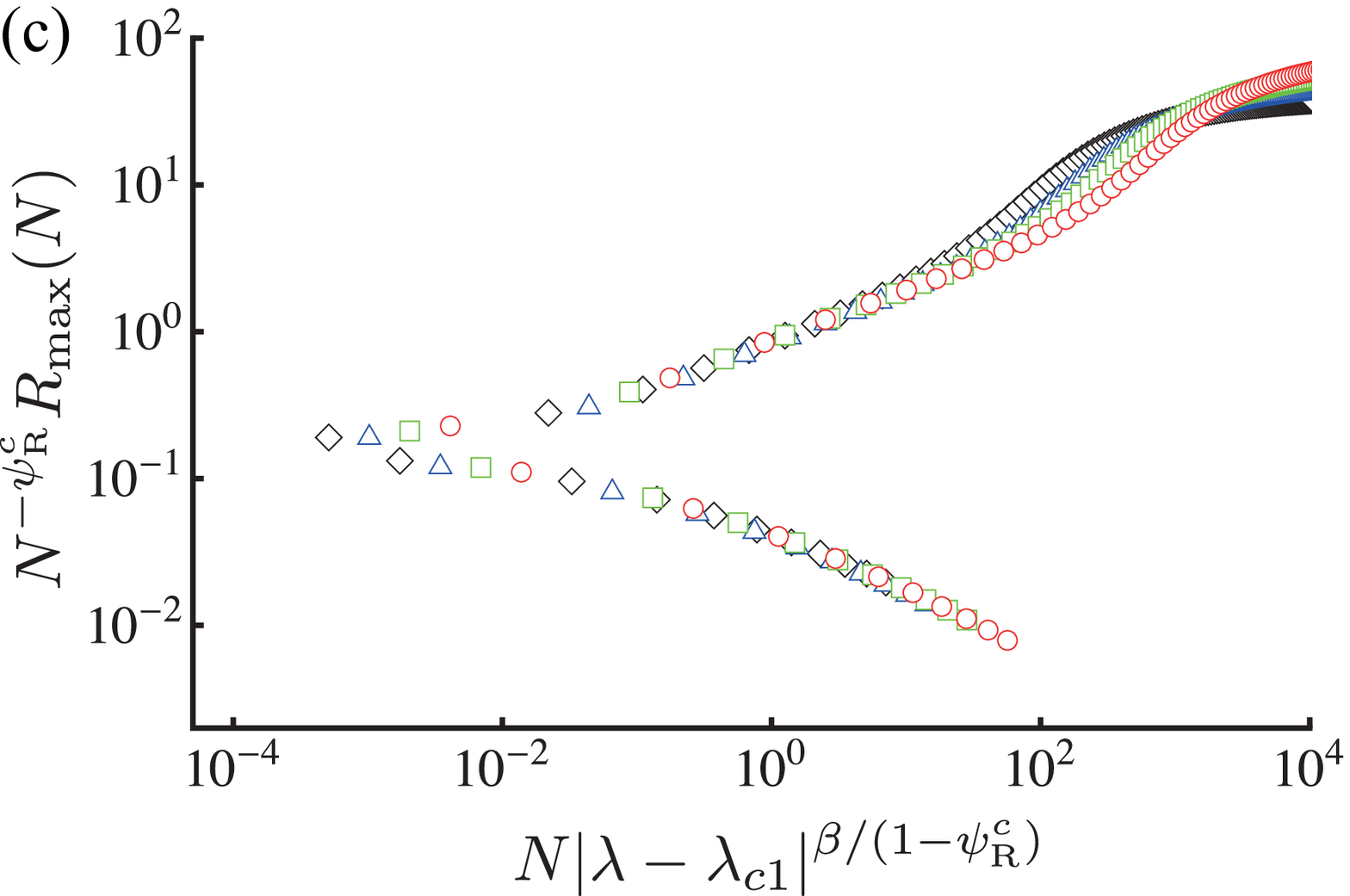}
   \end{center}
 \caption{
  (a) Fraction of the largest R-component $\rmax$ as a function of $\lambda$.
 Symbols were obtained from Monte-Carlo simulations.
 The solid line was drawn by evaluating Eq.~(\ref{eq:rmax}).
 (b) Numerical results of the fractal exponent of the largest R-component, $\pr$.
 Data with several values of $N$ have a crossing point at $\lcl \simeq 0.148$ and $\psi_{\rm R}^c \simeq 2/3$.
 (c) Finite size scaling for $R_{\rm max}$ around $\lcl$.
 Here we set $\psi_{\rm R}^c=2/3$ and $\beta=1$, 
 which means that the transition at $\lcl$ belongs to the mean-field universality class.
}
 \label{fig:rmax}
\end{figure}

\subsection{Derivation of $\lcl$}

To derive $\lcl$ for the SIR model with multiple seeds, we need to calculate the connectivity of the R-components generated by each seed. 
We should note that a percolation analysis, as in the previous subsection, using the degree distribution of the R-components $r_{\ell,0}/r$ 
is not applicable to this case, because such an analysis ignores the condition that {\it each R-component is connected}. 
To consider the connectivity of numerous R-components, we use the following procedure:
(i) We first calculate the probability, $P_n$, that the size of the R-component generated by a single seed is $n$.
(ii) For the case of $\rho>0$, the system has numerous R-components proportional to $\rho$ .
We regard each R-component as a supernode. 
The number of nodes confined in a supernode obeys the distribution $P_n$, and its degree $k_n$ is given accordingly.
(iii) Then, we consider a site percolation problem of supernodes. 
The first percolation point, $\lcl$, is given as the critical point where the infinitely connected component of the supernodes appears.

In  Appendix \ref{app:single}, we evaluate the mean size of the R-component starting from a single seed, 
$\langle n \rangle(=\sum_n P_n n)$, and the corresponding mean square size, $\langle n^2 \rangle=(\sum_n P_n n^2)$, by using generating functions.
Then, we consider the case of $\rho>0$.
Below $\lcl$, we naturally assume that the mean size $\langle n \rangle$ is so small that each R-component is a tree 
\footnote{Finite components have no cyclic path in infinite locally tree-like networks.} 
and that any overlaps between the R-components are negligible so that the total fraction of the R-components can be evaluated as $\rho \langle n \rangle$ 
\footnote{This condition means that the total size of the R-components is smaller than the number of nodes. Let us consider a finite network with $N$ nodes. 
The number of seeds is $N\rho$, and the mean R-component size for each seed is $\langle n \rangle$. 
Then the total size of the R-components is $N \rho \langle n \rangle$. 
Therefore, our assumption in the main text is equivalent to the statement that $N \rho \langle n \rangle<N$.}, 
which should necessarily be less than one. 
Then we can determine the creation process of the infinite R-component 
by regarding each R-component of size $n$ as a supernode whose degree depends on $n$ (see Fig.~\ref{fig:super}) 
and considering the percolation problem of these supernodes.

The density of susceptible nodes, $s$, is just one minus the density of the removed nodes, $\rho \langle n \rangle$:
\begin{equation}
s=1-\rho\langle n \rangle.
\end{equation}
Then the probability $\tilde{p}$ that the node reached by following a randomly chosen link is a component of a supernode is 
\begin{equation}
\tilde{p} = \frac{\rho \langle k_n \rangle}{\rho \langle k_n \rangle + z s}, \label{tildep}
\end{equation}
where $k_n$ is the number of external links of the R-component (the degree of the supernode) having size $n$, and is given by
\begin{equation}
k_n=(z-2) n +2. \label{kr}
\end{equation}
Equation (\ref{kr}) holds since each R-component is a tree with the number of edges equal to the number of nodes minus one.
The mean branching ratio of supernodes, $B$, is evaluated by multiplying $\tilde{p}$ by the mean excess degree of supernodes:
\begin{equation}
B=\tilde{p} \frac{\langle k_n(k_n-1)\rangle}{\langle k_n\rangle}=\frac{\rho\langle k_n(k_n-1)\rangle}{\rho \langle k_n\rangle + z s}.
\end{equation}
The percolation of supernodes takes place when $B \geq 1$, and thus the transition point is given by $B=1$.
That is, $\lcl$ satisfies 
\begin{align}
\frac{z}{\rho}&=\langle k_n(k_n-2)\rangle+z\langle n \rangle\notag\\
&=z_2^2\langle n^2\rangle +(3z-4)\langle  n \rangle, \label{eq:rhofunc}
\end{align}
where $\langle n \rangle$ and $\langle n^2 \rangle$ are functions of $\lambda$.

We can show that these moments diverge as $\langle n \rangle\sim(\lsir-\lambda)^{-1}$ and $\langle n^2 \rangle\sim(\lsir-\lambda)^{-3}$
when $\lambda$ approaches $\lsir$ from below (see Appendix \ref{app:single}), and therefore $\lcl\rightarrow\lsir$ as $\rho\rightarrow0$ like
\begin{equation}
\lsir-\lcl\sim \rho^{1/3}. \label{powerlaw-lambda}
\end{equation}
Thus, a small increase in $\rho$ drastically reduces $\lcl$ from $\lsir$.

We approximately have the fraction of the largest R-component, $\rmax$, 
by applying a procedure similar to the derivation of $\smax$ to the connected components of supernodes (see Appendix \ref{sec:component}).
This approximation may predict a rise of the order parameter $\rmax$ around $\lcl$, but inherently overestimates $\rmax$ for $\lambda > \lcl$ 
due to the overlaps between the R-components generated from each seed being non-negligible (see Fig.~\ref{fig:rmax} (a)).

We also check our estimate by comparison with Monte-Carlo simulations.
In Figs.~\ref{fig:rmax} (a) and (b), we plot the Monte-Carlo results for the order parameter, $\rmax(N)$, and the corresponding fractal exponent, 
$\pr(N) \equiv {\rm d}\ln \rmax(N)/{\rm d}\ln N$.
In a manner similar to that in the previous subsection, we find that the crossing point of $\pr$ is at our estimate of $\lcl$, $\lcl \simeq 0.148$.
We also find a good scaling result for $\Rmax(N)$ using $\pr^c \equiv \pr(\lcl)=2/3$, $\beta=1$, and the estimated $\lcl$ (Fig.~\ref{fig:rmax} (c)), 
supporting the validity of our estimate and indicating that the percolation transition at $\lcl$ belongs to the mean-field universality class.
 
\begin{figure}
 \begin{center}
  \includegraphics[width=70mm]{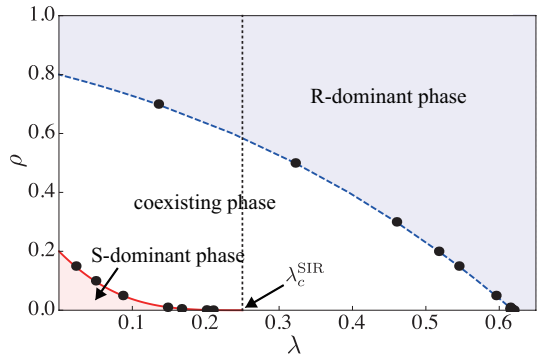}
   \end{center}
 \caption{
 Phase diagram in the $(\rho, \lambda)$-space. 
 Solid red and dashed blue lines are the analytically obtained $\lcl$ and $\lcu$, respectively.
 Black circles are plotted by the crossings of $\psi_{\rm R}$ and $\psi_{\rm S}$. 
}
 \label{fig:phase-diagram}
\end{figure}

\subsection{Phase diagram}

We analytically and numerically evaluate $\lcl$ and $\lcu$ for several values of $\rho$.
In Fig.~\ref{fig:phase-diagram}, we have the phase diagram of $(\rho, \lambda)$-space.
We find that our estimates of $\lcl$ and $\lcu$ perfectly match the Monte-Carlo results.
The first percolation point $\lcl$ is smaller than $\lsir$ as long as $\rho>0$. 
That is, the percolation of the R-components occurs without global outbreaks.
The gap between $\lcl$ and $\lsir$ shrinks with decreasing $\rho$ and $\lcl=\lsir$ in the limit $\rho \to 0$. 
Note that $\lcl=0$ when $\rho \ge 0.2$ because the seeds themselves percolate (the site percolation threshold of the $z$-RRG is $1/(z-1)$, which is derived from the local tree approximation \cite{newman2003structure,dorogovtsev2008critical}), 
and $\lcu=0$ when $\rho \ge 0.8$ because the seeds themselves disintegrate the susceptible network into finite components.

Our finite size scaling for several values of $\rho$ shows that 
both percolation transitions at $\lcl$ and $\lcu$ belong to the mean-field universality class, irrespective of the value of $\rho$. 
This seems unsurprising because the two processes comprising the present model, the SIR model and site percolation, 
belong to the mean-field universality class of percolation when the graph is RRG.

When $\rho > 0$, the system does not show any singular behavior at $\lsir$. 
However, this does not mean that $\lsir$ is unimportant. 
In practice, $\lsir$ is still an important measure in the strategy for disease control 
because a single seed has the potential to induce a global outbreak above $\lsir$ 
(in other words, the basic reproduction number $R_0>1$ when $\lambda > \lsir$). 
The singular behaviors at $\lcl$, e.g., the divergence of the R-susceptibility, may be interpreted as a precursor to global outbreaks, like the proverbial canary in a coal mine.

\begin{figure}
 \begin{center}
  \includegraphics[width=75mm]{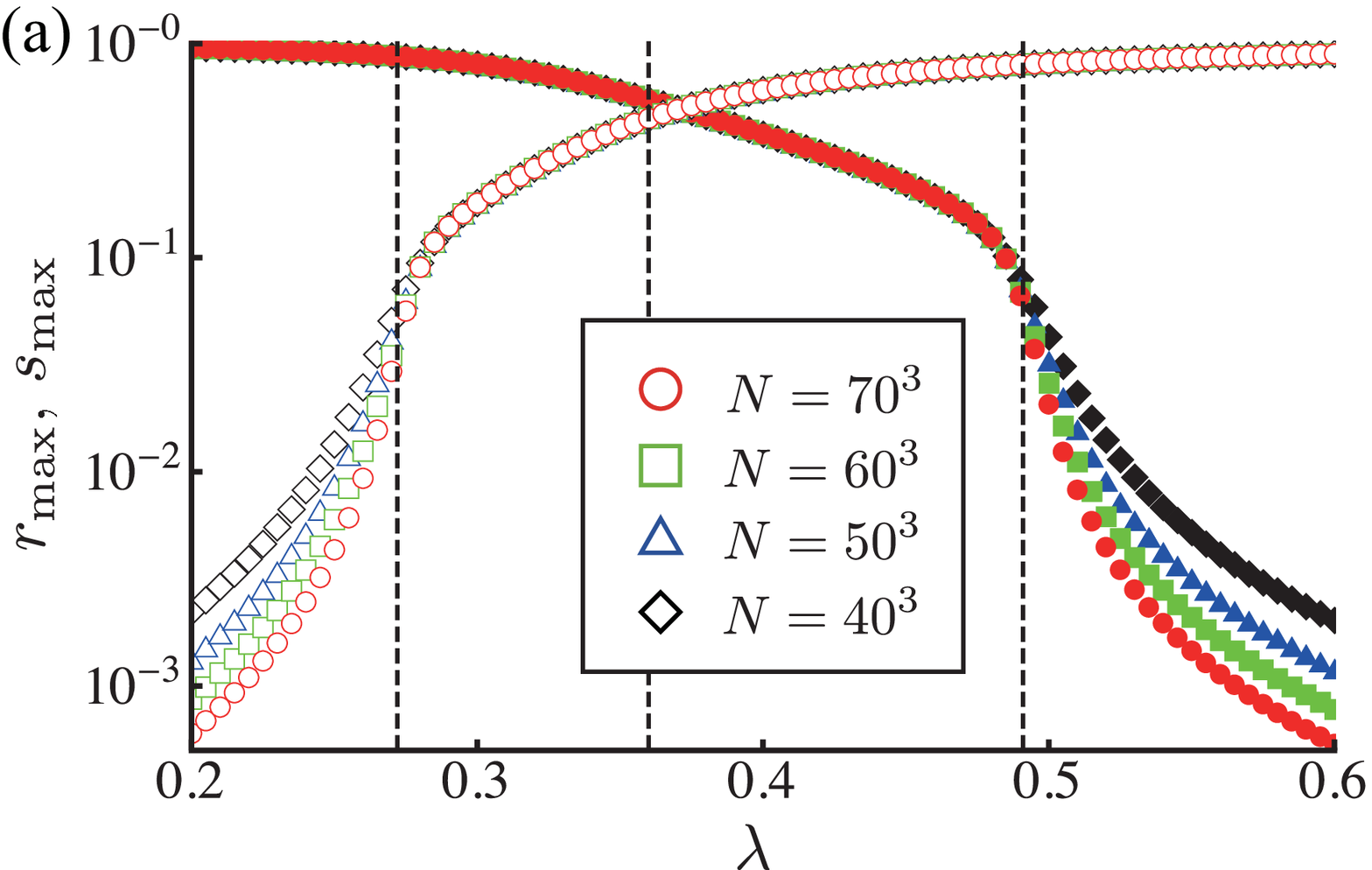}
  \includegraphics[width=75mm]{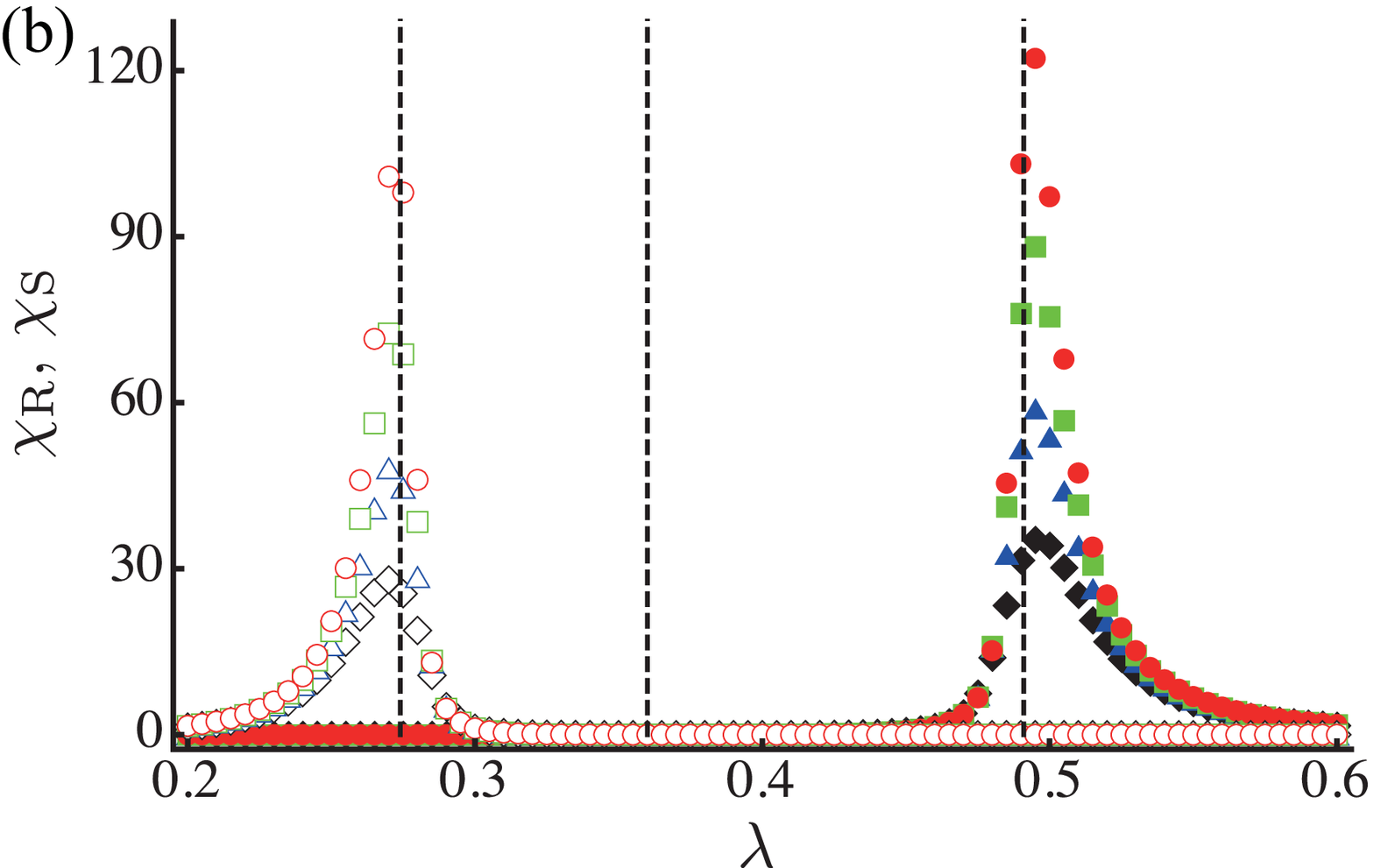}
   \end{center}
 \caption{
 Monte-Carlo results for the SIR model on a cubic lattice with the periodic boundary condition: 
 (a) fractions of the largest R-component (open symbols) and of the largest S-component  (filled symbols), $\rmax$ and $\smax$; (b) R-susceptibility (open symbols) and S-susceptibility (filled symbols), $\chi_{\rm R}$ and $\chi_{\rm S}$. 
 The three vertical lines represent $\lcl$, $\lsir$, and $\lcu$, from the left to the right. 
 Here $\lcl \simeq 0.27$ and $\lcu \simeq 0.49$ were confirmed 
 by the crossings of $\pr$ and $\ps$ and their finite size scalings, 
 and $\lsir \simeq 0.36$ was obtained from the Monte-Carlo simulations for the SIR model with a single seed (not shown).
}
 \label{fig:cubic}
\end{figure}

\begin{figure}
 \begin{center}
  \includegraphics[width=75mm]{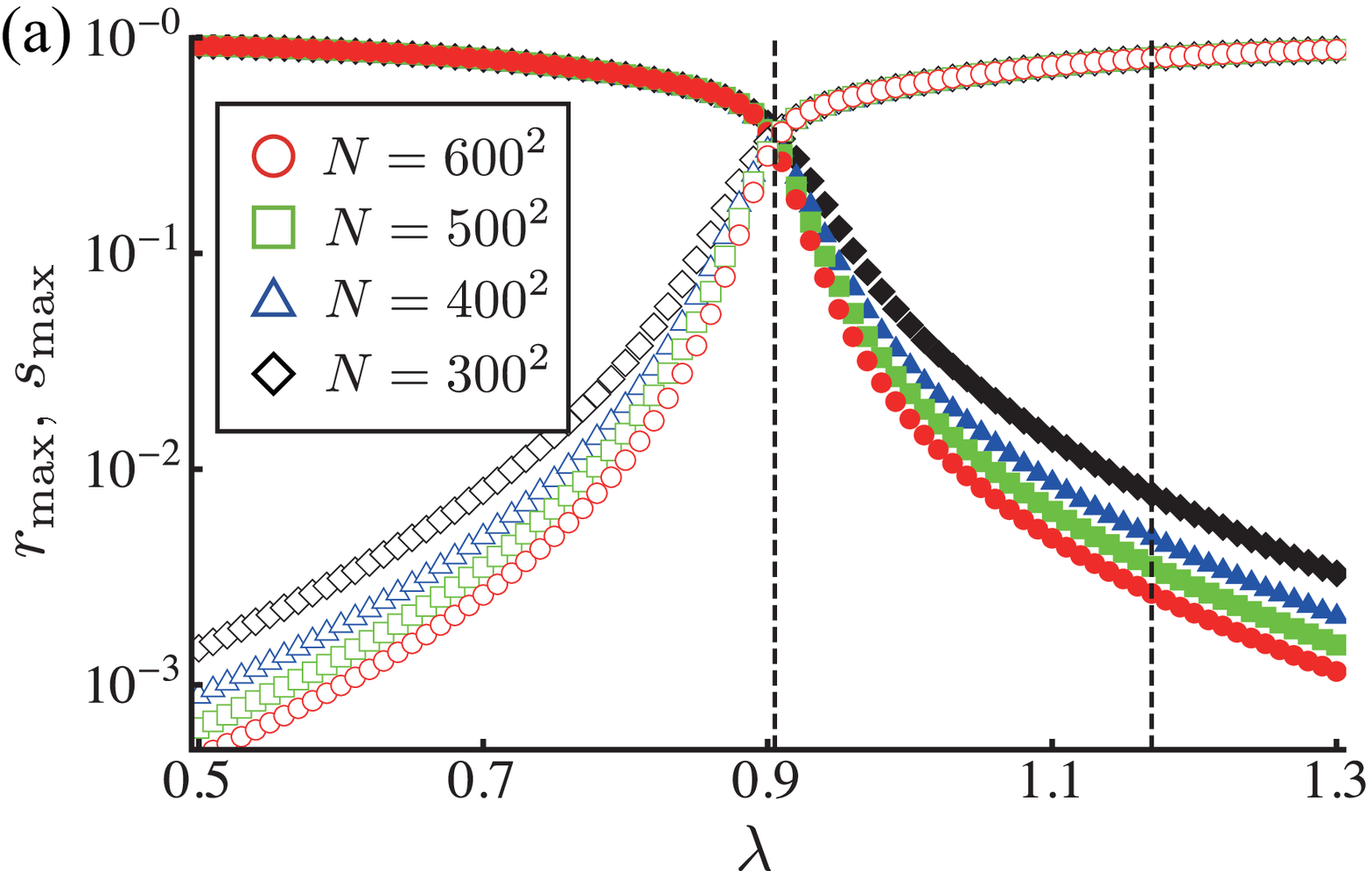}
  \includegraphics[width=75mm]{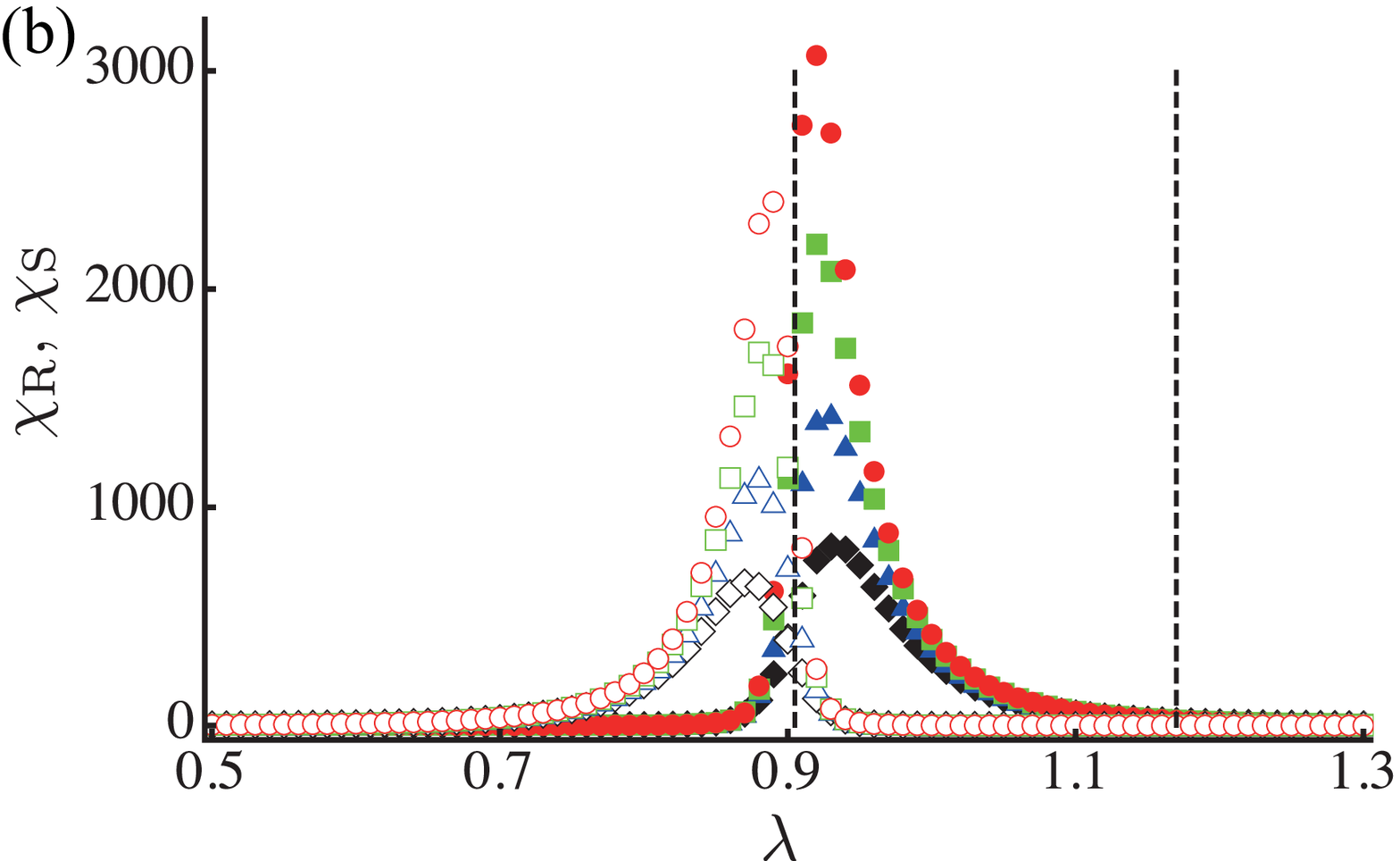}
   \end{center}
 \caption{
 Monte-Carlo results for the SIR model on a square lattice with the periodic boundary condition: 
 (a) fractions of the largest R-component (open symbols) and of the largest S-component (filled symbols), $\rmax$ and $\smax$; 
 (b) R-susceptibility (open symbols) and S-susceptibility (filled symbols), $\chi_{\rm R}$ and $\chi_{\rm S}$. 
 In these simulations, we set $\mu=1$ and $\rho=0.01$, $N=300^2, 400^2, 500^2, 600^2$. 
 Each of data is averaged over 50000 trials.
 The two vertical lines represent $\lcl=\lcu \simeq 0.91$ (left) and $\lsir \simeq 1.17$ (right), respectively.
 Here $\lcl=\lcu$ was confirmed by the crossings of $\pr$ and $\ps$ and their finite size scalings, 
 and $\lsir$ was obtained from the Monte-Carlo simulations for the SIR model with a single seed (not shown).
}
 \label{fig:square}
\end{figure}

\subsection{SIR model on regular lattices \label{spatial}}

We have numerically and analytically shown that the present model with multiple seeds on the RRG percolates 
at a lower infection rate than the epidemic threshold. 
Is this phenomenon in common with other networks, e.g., 
networks with many short loops and without the logarithmic dependence of the mean shortest path?
In the rest of this section, 
we briefly consider the SIR model with multiple seeds on finite dimensional Euclidean lattices by Monte-Carlo simulations. 

First, we consider the cubic lattice with the periodic boundary condition. 
In Monte-Carlo simulations, we set $\mu=1$, $\rho=0.01$, and $N=40^3, 50^3, 60^3, 70^3$. 
The number of trials is 50000.
In Fig.~\ref{fig:cubic} (a), we show the size dependence of the order parameters, $\rmax$ and $\smax$. 
We find that $\rmax$ and $\smax$ become nonzero and zero 
at different points $\lcl \simeq 0.27$ and $\lcu \simeq 0.49$, respectively. 
This discrepancy is also led by the peak positions of R-susceptibility and S-susceptibility, as shown in Fig.~\ref{fig:cubic} (b). 
Both $\lcl$ and $\lcu$ are also different from the epidemic threshold $\lsir$, 
which is obtained from the single seed simulations (not shown). 
Starting from a single seed, $\rmax(=r)$ becomes nonzero at $\lsir \simeq 0.36$, which is larger than $\lcl$.
Then, we have the discrepancy among $\lcl, \lsir$, and $\lcu$, on the cubic lattice, 
and expect that the observed phenomena on the RRG will hold for other clustered networks.
A qualitative difference between the cubic lattice and the RRG is on their universality class. 
From the crossings of the fractal exponents $\pr$ and $\ps$, 
we have $\pr^c \simeq \ps^c \simeq 0.84$ for the cubic lattice (not shown), and 
this estimate means that transitions of both R-components and S-components belong to 
the universality class of three-dimensional percolation transition \cite{stauffer1994introduction}, 
but not to the mean-field universality class.

Let us mention a special case, the SIR model in the square lattice with the periodic boundary condition.
In Fig.~\ref{fig:square}, we show the Monte-Carlo results.
In this case, $\smax$ and $\rmax$ seem to undergo a macroscopic change 
at the same point, i.e., $\lcl=\lcu \simeq 0.91$ (see Fig.~\ref{fig:square} (a)). 
The corresponding susceptibilities also seem to have a peak at the same value of $\lambda$ (Fig.~\ref{fig:square} (b)). 
Compared to single seed simulations, $\lcl$ and $\lcu$ decrease with an increase in $\rho$ 
and differ from the epidemic threshold $\lsir \simeq 1.17$.
The absence of the coexisting phase does not seem surprising if we consider a spatial constraint of the square lattice:  
When the largest component percolates the lattice vertically and horizontally, 
the residual components after removing the largest one cannot maintain the connection across the lattice. 
(This reflects on the fact that the percolation threshold is equal to or larger than 1/2 in both site and bond percolations.)
Turning to other real spatial networks, which are often regarded as two-dimensional objects, 
it is an open question whether or not the coexisting phase exists. 


\section{Summary}

In this paper, we have studied the SIR model in an RRG with a nontrivial fraction of infection seeds, $\rho$. 
Through analytical estimates and numerical simulations, we have obtained the phase diagram in ($\rho,\lambda$)-space.
The SIR model with numerous seeds shows the percolation transition of the removed and susceptible nodes at $\lcl$ and $\lcu$, respectively.
In particular, $\lcl$ is smaller than the epidemic threshold $\lsir$ as long as $\rho>0$.
This means that epidemic clusters generated by multiple seeds percolate without global outbreaks.

So far, we have focused on the SIR model on the RRG and the lattices. 
We expect that the above statement holds for the SIR model on other networks 
although the details of the phase transition may depend on network structures, e.g., $\lcl<\lsir=0$ in a fat-tailed scale-free network with $\gamma \le 3$.

Finally, we briefly discuss other epidemic models with multiple seeds. 
Krapivsky et al. \cite{krapivsky2011reinforcement} proposed an extended SIR model, called a transient fad, with the assumption of a well-mixed population.
They analytically showed that this model exhibits a discontinuous transition if $\rho>0$. 
The authors and a collaborator \footnote{T.\,Hasegawa, N.\,Kinoshita, and K.\,Nemoto, in preparation.} 
performed Monte-Carlo simulations for this fad model in networks to confirm that 
a discontinuous jump of the order parameter appears near the epidemic threshold, which is behind the percolation of epidemic clusters. 
The authors also investigated the discrete time version of the transient fad to confirm it numerically and analytically \cite{hasegawa2014discontinuous}.
Very recently, several generalized epidemic models on networks beyond the classical SIR model have been investigated \cite{bizhani2012discontinuous,chung2014generalized,cai2015avalanche,gomez2015abrupt}.
It will be interesting to clarify what numerous seeds induce in such generalized epidemic models. 

\section*{Acknowledgements}
T.H.'s work was partially supported by Grant-in-Aid for Young Scientists (B) (No.~24740054 and No.~15K17716) 
and Grant-in-Aid for Scientific Research (B) (No.~26310203) from the Japan Society for the Promotion of Science (JSPS), and by JST, ERATO, Kawarabayashi Large Graph Project.


\appendix 

\section{Derivation of $\lsir$ for the limit $\rho \to 0$ \label{sec:derivation-lsir}}

The AMEs do not predict any transition point when $\rho>0$ because $r>\rho>0$. 
However, it is possible to derive the transition point $\lsir$ if the fraction of infection seeds is infinitesimally small ($\rho \to 0$).
In the AMEs for $\{s_{l,m}\}$, only $s_{z,0}$ and $s_{z-1,1}$ are relevant up to the first order of $\rho$,
\begin{equation}
\dot s_{z,0} = -\beta_s^{si}zs_{z,0},\quad 
\dot s_{z-1,1} 
= -\lambda s_{z-1,1} + \beta_s^{si}zs_{z,0} - \beta_s^{ir}s_{z-1,1},
 \label{AME-eq-rho0}
\end{equation}
with 
\begin{equation}
\beta_s^{si}=\lambda\frac{(z-1)s_{z-1,1}}{zs_{z,0}},\quad \beta_s^{ir}=\mu, \label{AME-rate-rho0}
\end{equation}
and the initial condition
\begin{equation}
s_{z,0}(0)=1-(z+1)\rho,\quad s_{z-1,1}(0)=z\rho. 
\end{equation}
Equation (\ref{AME-rate-rho0}) is substituted into Eq.~(\ref{AME-eq-rho0}) to find the condition for $s_{z-1,1}$ to remain finite as $t\to\infty$,
\begin{equation}
\frac{\mu}{\lambda}>z-2,
\end{equation}
and thus, the lower bound of $\mu/\lambda$ gives the epidemic threshold, $\lsir=\mu/(z-2)$, which corresponds to the known result (\ref{lambda_sir}).

\begin{figure}
 \begin{center}
  \includegraphics[width=100mm]{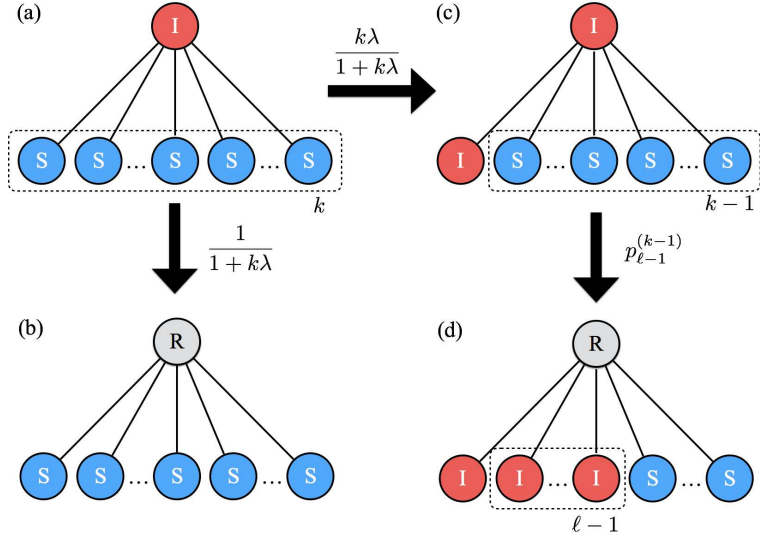}
   \end{center}
 \caption{
Schematic of the recursive relation for $p_{\ell}^{(k)}$. 
Let consider the probability $p_\ell^{(k)}$ that an infected node infects $\ell$ nodes of $k$ susceptible neighbors 
before being removed [(a)$\to$(d)]. 
The probability that this infected node is removed 
before infecting any neighbors [(a)$\to$(b)] is $p_0^{(k)}=1/(1+k \lambda)$. 
On the other hand, 
the event that this node infects one susceptible neighbor [(a)$\to$(c)] occurs 
with probability $1-p_0^{(k)}=k\lambda/(1+k \lambda)$. 
At (c), the focal infected node infects further $\ell-1$ nodes from $k-1$ remaining neighbors 
before being removed [(c)$\to$(d)] with probability $p_{\ell-1}^{(k-1)}$ because the infecting process is Markovian. 
Thus, $p_{\ell}^{(k)}$, which is the probability from (a) to (d), is given as $p_{\ell-1}^{(k-1)} \times k\lambda/(1+k\lambda)$.
}
 \label{fig:p_l^k}
\end{figure}

\section{Distribution of the R-component generated from a single seed \label{app:single}}

First, we evaluate the size distribution of the R-component created by a single seed.
To do this, we need to know the probability, $p_\ell^{(k)}$, that an infected node will infect $\ell$ of $k$ neighboring susceptible nodes before being removed.
Such an infected node is removed before infecting any neighbors with a probability $p_0^{(k)}=\frac{1}{1+k \lambda}$,
so that the probability of its infecting at least one neighboring node before removal is $1-p_0^{(k)}=\frac{k\lambda}{1+k\lambda}$.
As shown in Fig.~\ref{fig:p_l^k}, we can express $p_\ell^{(k)}$ as a recursive form,
\begin{equation}
p_\ell^{(k)}=\begin{cases}
\frac{1}{1+k\lambda},&\ell=0,\\
\frac{k\lambda}{1+k\lambda}\times p_{\ell-1}^{(k-1)},& 0<\ell\le k.
\end{cases}\label{eq:gz}
\end{equation}
From this equation, we find that the generating function of $p_\ell^{(k)}$,
\begin{equation}
g_k(x)=\sum_{\ell=0}^k p_\ell^{(k)} x^\ell,
\end{equation}
satisfies the recursion relation
\begin{equation}
g_k(x)=\frac{1+k\lambda x g_{k-1}(x)}{1+k\lambda}
\end{equation}
with $g_0(x)=1$.

\begin{figure}
 \begin{center}
  \includegraphics[width=150mm]{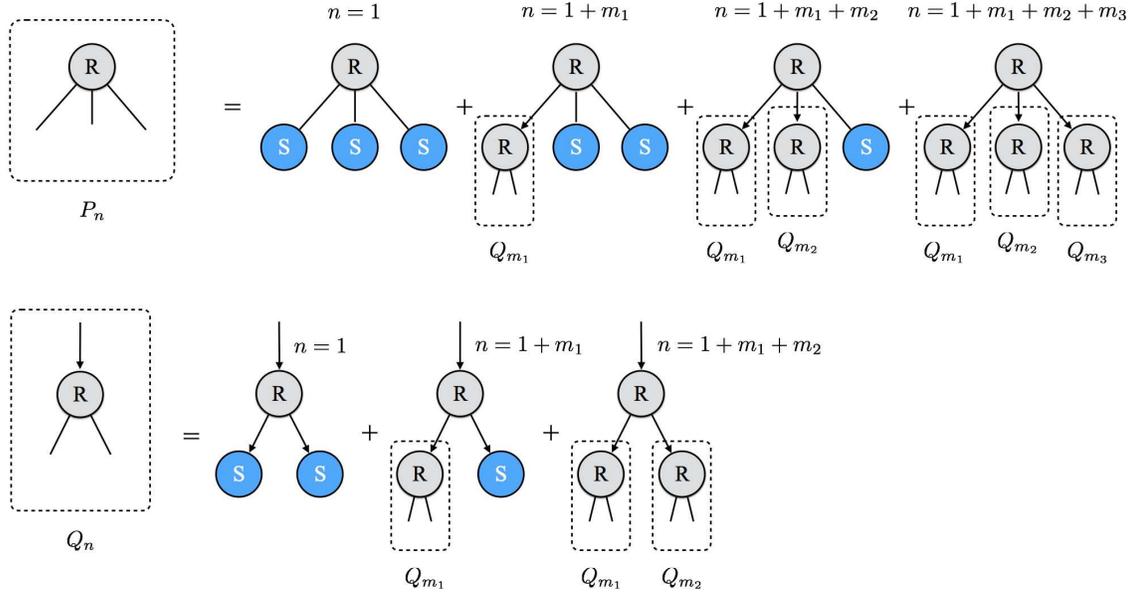}
   \end{center}
 \caption{
Schematic of $P_n$ (top) and $Q_n$ (bottom) for the case of the RRG with $z=3$: 
$P_n=p_0^{(3)}\delta_{n,1}+p_1^{(3)} \sum_{m_1} Q_{m_1} \delta_{n,1+m_1}+p_2^{(3)} \sum_{m_1,m_2} Q_{m_1}Q_{m_2} \delta_{n,1+m_1+m_2}+p_3^{(3)} \sum_{m_1,m_2,m_3} Q_{m_1} Q_{m_2} Q_{m_3} \delta_{n,1+m_1+m_2+m_3}$, 
and $Q_n=p_0^{(2)}\delta_{n,1}+p_1^{(2)} \sum_{m_1} Q_{m_1} \delta_{n,1+m_1}+p_2^{(2)} \sum_{m_1,m_2} Q_{m_1}Q_{m_2} \delta_{n,1+m_1+m_2}$.
}
 \label{fig:PandQ}
\end{figure}

Now let $P_n$ be the probability that a single seed creates an R-component of size $n$,
and let $Q_n$ be the probability that a node infected by another node further creates a partial R-component of size $n$.
Then, by considering the infection process starting from a single seed, $P_n$ can be evaluated  as
\begin{equation}
P_n=\sum_{\ell=0}^{z}p_\ell^{(z)}\sum_{m_1,m_2,\cdots,m_\ell}Q_{m_1}Q_{m_2}\cdots Q_{m_\ell}
\delta_{n,1+\sum_{\mu=0}^\ell m_\mu},\label{eq:Pn}
\end{equation}
and similarly, $Q_n$ can be recursively given as
\begin{equation}
Q_n=\sum_{\ell=0}^{z_1}p_\ell^{(z_1)}\sum_{m_1,m_2,\cdots,m_\ell}Q_{m_1}Q_{m_2}\cdots Q_{m_\ell}
\delta_{n,1+\sum_{\mu=0}^\ell m_\mu},\label{eq:Qn}
\end{equation}
where $z_\nu=z-\nu$ (Fig.~\ref{fig:PandQ}). Introducing the corresponding generating functions,
\begin{equation}
G_0(x)=\sum_{n=1}^\infty P_n x^n,\quad G_1(x)=\sum_{n=1}^\infty Q_n x^n,
\end{equation}
we can express the above relations (\ref{eq:Pn}) and (\ref{eq:Qn})  in a compact form as
\begin{equation}
G_0(x)=xg_z(G_1(x))
\end{equation}
and
\begin{equation}
G_1(x)=xg_{z_1}(G_1(x)),
\end{equation}
respectively.

What we want to know is the mean size of the R-component, $\langle n \rangle= G_0'(1)$, 
and the mean square size of the R-component, $\langle n^2 \rangle=G_0''(1)+G_0'(1)$.
To evaluate these values, we need the derivative of $g_k(x)$, which is given by
\begin{equation}
g_k'(x)=\frac{k\lambda}{1+k\lambda}[g_{k-1}(x)+xg_{k-1}'(x)],
\end{equation}
from which the mean value $\langle \ell \rangle_k$ can be easily found to be
\begin{equation}
\langle \ell \rangle_k = g_k'(1)=\frac{k\lambda}{1+\lambda}.
\end{equation}
One also requires the second derivative,
\begin{equation}
g_k''(x)=\frac{k\lambda}{1+k\lambda}[2g_{k-1}'(x)+xg_{k-1}''(x)],
\end{equation}
so that 
\begin{equation}
\langle\ell(\ell-1)\rangle_k=g_k''(1)=\frac{2k(k-1)\lambda^2}{(1+\lambda)(1+2\lambda)},
\end{equation}
yielding
\begin{equation}
\langle\delta\ell^2\rangle_k=
\langle\ell^2\rangle_k-\langle\ell\rangle_k^2=\frac{k\lambda[1+(k+1)\lambda]}{(1+\lambda)^2(1+2\lambda)}.
\end{equation}

Now we can evaluate the derivatives of $G_0(x)$ and $G_1(x)$ as
\begin{equation}
G_0'=g_z(G_1)+xg_z'(G_1)G_1',
\end{equation}
\begin{equation}
G_0''=2g_z'(G_1)G_1'+xg_z''(G_1)G_1'^2+xg_z'(G_1)G_1'',
\end{equation}
and
\begin{equation}
G_1'=g_{z_1}(G_1)+xg_{z_1}'(G_1)G_1',
\end{equation}
\begin{equation}
G_1''=2g_{z_1}'(G_1)G_1'+xg_{z_1}''(G_1)G_1'^2+xg_{z_1}'(G_1)G_1''.
\end{equation}
Setting $x=1$ gives $G_1(1)=1$ and
\begin{equation}
G_1'(1)=1+g_{z_1}'(1)G_1'(1)=\frac{1}{1-g_{z_1}'(1)}=\frac{1}{1-\langle\ell\rangle_{z_1}},\label{eq:G1d1}
\end{equation}
\begin{align}
G_1''(1) &=2g_{z_1}'(1)G_1'(1)+g_{z_1}''(1)G_1'(1)^2+g_{z_1}'(1)G_1''(1)\notag\\
&=G_1'(1)^3[2g_{z_1}'(1)(1-g_{z_1}'(1))+g_{z_1}''(1)].
\end{align}
These quantities provide an explicit expression for 
the first moment, $M_1$, and the second cumulant, $C_2$ (and thus the second moment $M_2=C_2+M_1^2$), of $P_n$ as
\begin{equation}
M_1=\langle n \rangle=G_0'(1)=1+\langle\ell\rangle_z G_1'(1),
\end{equation}
\begin{align}
C_2&=\langle n^2\rangle - \langle n \rangle^2\notag \\
&=G_0''(1)+M_1-M_1^2\notag\\
&=G_1'(1)^2[g_z''(1)+g_z'(1)-g_z'(1)^2]+g_z'(1)[G_1''(1)+G_1'(1)-G_1'(1)^2]\notag\\
&=G_1'(1)^2\langle\delta\ell^2\rangle_z+G_1'(1)^3\langle\ell\rangle_z \langle\delta\ell^2\rangle_{z_1}.
\end{align}
When $\lambda$ approaches $\lsir$ from below, $G_1'(1)$ dominates the behavior of $M_1$ and $C_2$.
Indeed (\ref{eq:G1d1}) tells us that $G_1'(1)$ diverges as $\delta\lambda^{-1}$, where $\delta\lambda=\lsir-\lambda$, and thus
\begin{equation}
M_1=\langle n\rangle\sim\delta\lambda^{-1},\quad
C_2\sim\langle n^2\rangle\sim\delta\lambda^{-3}. \label{eq:M1andC2}
\end{equation}
Substituting Eq.~(\ref{eq:M1andC2}) for Eq.~(\ref{eq:rhofunc}), 
we have a power-law dependence of the gap $\delta \lambda$ on the seed fraction $\rho$ as Eq.~(\ref{powerlaw-lambda}).

\section{Largest R-component size \label{sec:component}}

The generating function of the probability that a supernode will have the degree $k_n$ is given by
\begin{equation}
H_0(x)=\sum_{n=1}^\infty P_n x^{k_n}=x^2G_0(x^{z_2}),
\end{equation}
and that of the excess degree as
\begin{equation}
H_1(x)=\frac{H_0'(x)}{H_0'(1)}=\frac{2xG_0(x^{z_2})+z_2x^{z_1}G_0'(x^{z_2})}{\langle k_n \rangle}.
\end{equation}
Let $u$ be the probability that a finite cluster of supernodes is found by following randomly chosen links; this satisfies
\begin{equation}
u=1-\tilde{p} + \tilde{p} H_1(u).
\end{equation}
Here, $\tilde{p}$ is the probability that the node reached by following a randomly chosen link is a component of a supernode and is given by Eq.~(\ref{tildep}). 
Then, the density of the largest component of supernodes in size, i.e., $\rmax$, is evaluated as
\begin{equation}
\rmax=\rho\sum_{n=1}^\infty n P_n(1-u^{k_n})=\rho[\langle n \rangle -u^zG_0'(u^{z_2})], \label{eq:rmax}
\end{equation}
where we have used $\langle k_n \rangle = (z-2) \langle n \rangle +2$ from Eq.~(\ref{kr}).

\end{document}